\documentclass[10pt]{article}
\usepackage[top=0.85in,footskip=0.5in]{geometry}

\usepackage{amsmath,amssymb}

\usepackage[utf8x]{inputenc}

\usepackage{cite}

\usepackage[aboveskip=1pt,labelfont=bf,labelsep=period,singlelinecheck=off]{caption}

\usepackage{lastpage,fancyhdr,graphicx}
\usepackage{epstopdf}

\begin{document}
\vspace*{0.05in}

\begin{flushleft}
{\Large
\textbf\newline{Preliminary experiments demonstrating a ``directed" Maxwell's granular demon} 
}
\newline

Frank Corrales\textsuperscript{1},
Yuri Nahmad\textsuperscript{2*},
Ernesto Altshuler\textsuperscript{1*}
\\
\bigskip
\textbf{1} Group of Complex Systems and Statistical Physics, University of Havana, 10400 Habana, Cuba
\\
\textbf{2} Instituto de Física, Zona Universitaria San Luis Potosí, SLP 78290, México
\\
\bigskip

* ealtshuler@fisica.uh.cu
\newline
* yuri@ifisica.uaslp.mx

\end{flushleft}

\section*{Abstract}
In this paper, we design a system of two symmetrical containers communicated by an
aperture, in which a granular gas of glass spheres is created by shaking laterally the whole
system in a planetary mill. If the aperture consists in a symmetrical hole, the
two halves end up with the same number of grains after some time when initially all particles
are into in one of the containers. However, when a funnel-like aperture is used, a robust symmetry breaking is induced:
if all the grains are originally deposited in the container facing the wide side $95 \%$ of the grains
pass to the opposite side in a relatively small time.

\section*{Introduction}

Granular matter in motion is known to display many unexpected behaviors:
``Brazil-nuts" effects \cite{knight1993vibration} and wave patterns \cite{melo1994transition} in vibrated grains;
ticking sand in hourglasses \cite{le1996ticking},
avalanches and uphill solitary waves in granular flows \cite{altshuler2008revolving,martinez2007uphill},
``revolving rivers" in conical sandpiles \cite{altshuler2003sandpile}, gravity-independent penetration of 
intruders in granular beds \cite{altshuler2014settling}
and many others \cite{kadanoff1999built}.

We concentrate here on one of the most puzzling granular phenomena:
dense versus dilute phase separation into indistinguishable and equivalent 
communicated compartments, spontaneously occurs in vibrated granular systems.

It was first observed by Schlichting and Nordmeier in 1996 \cite{Nordmeier1996} and then
approached experimentally or theoretically by others
\cite{eggers1999sand,brey2001hydrodynamic,van2001hysteretic,evesque2002phase,isert2009influence,opsomer2013dynamical}.
Such effects have been generically called {\it granular Maxwell demons}.
One of their common features is the {\it spontaneity} of the specific
direction of the symmetry breaking; the formation of the cluster takes place
randomly either in the left or in the right containers.

In granular demons the symmetry breaking is believed to occur due to the dissipative collisions between grains.
Thanks to that, if a set of colliding particles locally decreases
the gas ``temperature" (i.e., the average kinetic energy of the particles), new grains
collide with those and cool down, resulting in the amplification of the fluctuation:
a large cluster (or clusters) \cite{falcon1999cluster,isert2009influence} can be eventually formed into one of the 
connected containers, then breaking the symmetry.

In this paper, we ``catalyze" --so to speak-- the classical two-compartment granular Maxwell demon: by using a funnel-like
geometry of the aperture between the two spaces containing a granular gas, we promote a
symmetry breaking in their densities even under the same experimental conditions where it does not arise using a
conventional, symmetric aperture. Moreover, we are able to control (or direct) in what direction the symmetry
will be broken.

\section*{Materials and methods}

Our experimental setup consists in a cubic container of $10$ cm-side divided into two equal volumes by a vertical wall
along the diagonal direction. The two volumes are communicated by a funnel-like channel of $14$ mm height, with
large and small holes of diameters $21$ and $4$ mm, respectively, as shown in ``Fig. 1(a)''. Two cubical boxes like that are mounted
into a planetary mill, describing the motion sketched in ``Fig. 1(b)''.

\begin{figure}[!h]
\centering      \includegraphics[width=3.5in, height=2in]{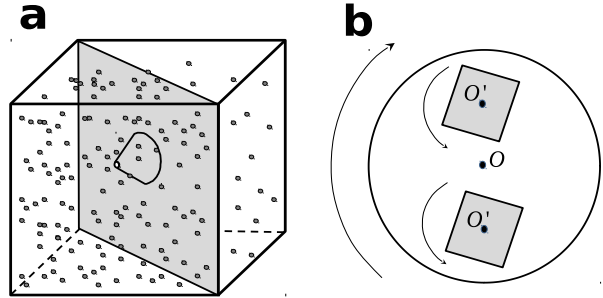}
  \caption{{\bf The experimental setup.} (a) Glass cube split into two equal spaces using a funnel-like aperture. 
  The small particles are $2$ mm diameter glass beads. (b) Top view of the planetary shaker where two gas containers 
  are simultaneously studied (The main platform revolves clockwise around the vertical axis $O$ at $1$ Hz, while each 
  container revolves counterclockwise around $O'$ at $0.75$ Hz, corresponding to the ```rotation 3" in the planetary mill's 
  rotation indicator). The curved arrows indicate the rotational motion of the different parts of the device.}
\label{fig:setup}
\end{figure}

As a result, the system experiences a strong agitation. In a typical experiment, $10 000$ glass beads (each one with a
mass of $0.01$ g)  are introduced into the volume corresponding to the wide side of the funnel aperture (Space 1 in ``Fig. 2'').
Due to the strong agitation, 
they behave like a granular gas. ``Fig. 2'' shows snapshots taken from top, at different moments, where the total
of $10 000$ grains occupied space 1.

 \begin{figure}[!h]
\centering      \includegraphics[width=4.5in, height=2.5in]{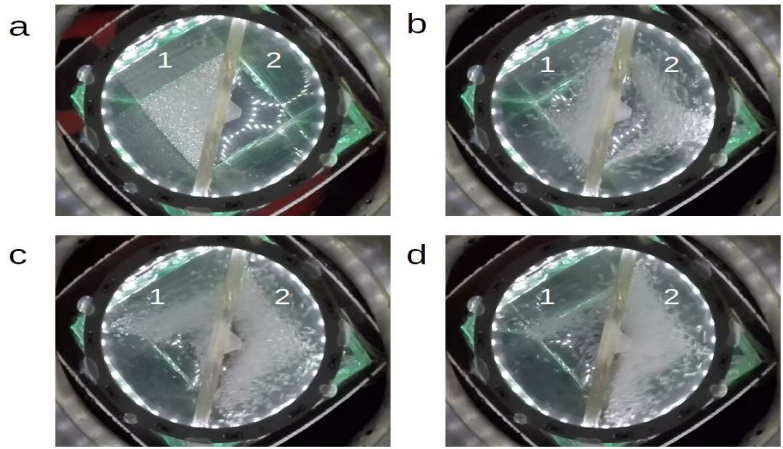}
 \caption{ {\bf Directing a Maxwell granular demon.} Snapshots taken from the top of one of the boxes at
 (a) before starting the experiment, and (b) $200$, (c) $500$ and (d) $3000$ seconds after the beginning of the
 experiment, which started with a total of $10000$ beads into semi-space 1. The bright spots are spurious reflections 
 associated to the light emitting diodes used to illuminate the photographs.}
 \label{fig:setup}
\end{figure}

In order to quantify in detail the temporal evolution of the particle distribution, we performed experiments
where the planetary mill was turned off every few seconds after the beginning of the experiment,  and the number of
beads into the originally empty space was determined by weighting them. Then, the beads were put back and the agitation 
was re-started.

\section*{Results}

The two upper curves in ``Fig. 3'' show the temporal evolution of the number of glass beads in the half-container
that was originally empty (each curve correspond to one of two identical boxed mounted on the planetary mill).
As can be seen, nearly $95 \%$ of the beads have passed to the opposite space after $2000$ seconds.
The inset illustrates that, the time evolution approaches a power law.

\begin{figure}[!h]
\centering      \includegraphics[width=5.8in, height=3.8in]{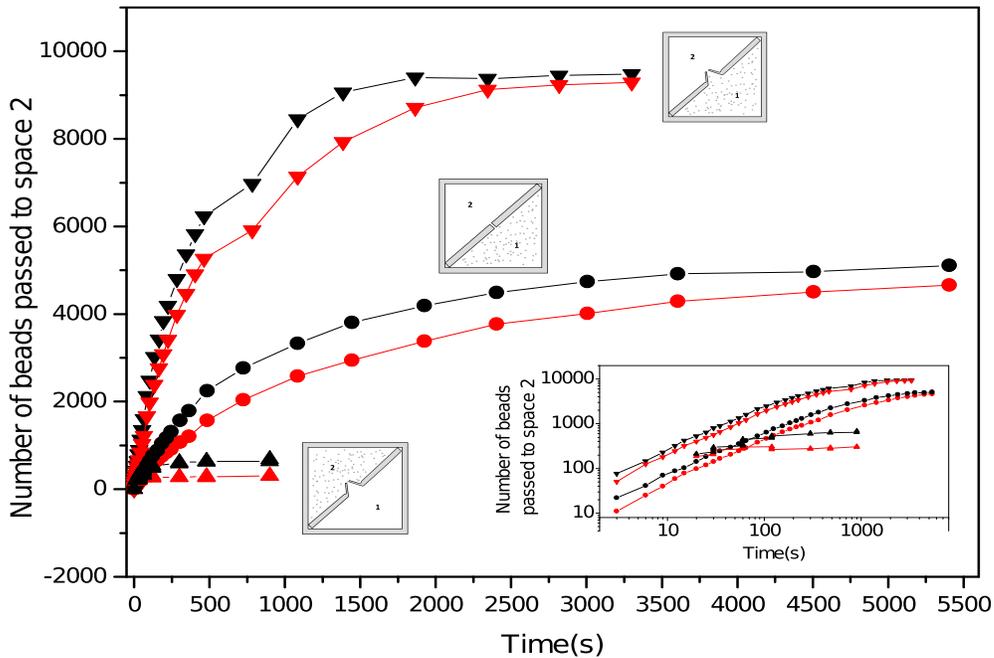}
 \caption{ {\bf Time evolution of the number of particles for three different types of barriers.} 
 The curves correspond to the filling up of one half of the cube due to the passage of beads from the opposite side,
 that contained the 10 000 available beads at the beginning of the experiment. Each one of the three pairs of curves correspond 
 to the initial conditions sketched near each of them. The pairs of curves for each case are the results of two
 repetitions of each type of experiment.  (Notice that the experimental situation 
 illustrated in ``Figs 1 and 2'' is that corresponding to the two upper curves here).}
 \label{fig:setup}
\end{figure}

The bottom curves correspond to a similar experiment, except for the fact that the $10000$ beads have been originally 
situated in the container facing the small aperture of the funnel. As can be seen, only $5 \%$ of the beads have managed
to pass to the opposite space, in good agreement with the previous experiment.

Finally, the pair of curves shown in the middle of the graph correspond to an aperture consisting in a simple hole of $4$ mm
diameter. At the beginning of the experiment, the $10 000$ beads were deposited on one of the halves. After $6500$ seconds
(not shown in the graph of ``Fig. 3''), the
two half-spaces contain $50 \%$ of the beads: the symmetry has not been broken. However, the inset shows a power law
relaxation in the number of beads, just as in the case of the first experiment. It is worth noting that, if this experiment
is started with $50 \%$ of the beads into each half --i.e., the common practice in analogous experiments \cite{Nordmeier1996}--
the final proportions are maintained, within small fluctuations. Moreover,
the symmetry is not broken even for smaller agitation strengths, a somewhat unexpected result in the light of Egger's model
\cite{eggers1999sand}.

\section*{Conclusion}

With the introduction of the funnel-like aperture we have ``catalyzed'' the Maxwell demon effect in a system
where it does not appear using a symmetric connection between containers. We produce the effect even when the system
starts from the opposite symmetry --i.e., the container initially full is emptied to fill up the opposite one.
Our system is closer to the original Maxwell's idea than usual experiments, in the sense that the symmetry breaking action is triggered by a ``selective" element 
(the funnel-like aperture), instead of depending on the random formation of clusters in {\it any} of the
two symmetrical containers. The funnel also increases the evolution speed of the system: while the symmetry is
reversed within $2000$ seconds, the substitution of the funnel by a symmetrical hole takes more than twice as much time to reach
the symmetric state (see ``Fig. 3'', top and middle curves, respectively).
Even when our aperture is very small compared to the separating wall (its area is just a $3\%$ of the total wall area), 
the presence of the funnel introduces a ``ratchet effect" capable of changing the overall dynamics of the granular system in a 
relatively small time. In fact, related effects have been observed at a microscopic scale in silicon membranes with asymmetric pores,
when acting as ``massive parallel brownian ratchets'' \cite{matthias2003asymmetric}, and in millimetric disks, as they diffuse
through a membrane made of asymmetric pores \cite{shaw2007geometry}.

\section*{Supporting information}

\paragraph*{S1 File.}
\label{S1_File}
{\bf Experimental data corresponding to ``Fig. 3''. (PDF)

\paragraph*{S2 Video.}
\label{S2_Video}
{\bf Sample video corresponding to ``Fig. 2b''.}

\end{document}